\documentclass[aps,prl,twocolumn,superscriptaddress,showpacs]{revtex4-1}
\usepackage{graphicx}
\usepackage{calc}
\usepackage{bm}

\begin{document}

\title{Quantum Hall Mach-Zehnder interferometer at fractional filling factors}

\author{E.V.~Deviatov}
\affiliation{Institute of Solid State Physics RAS, Chernogolovka, Moscow District, 142432, Russia}
\affiliation{Moscow Institute of Physics and Technology, Institutsky per. 9, Dolgoprudny, 141700 Russia}

\author{S.V.~Egorov}
\affiliation{Institute of Solid State Physics RAS, Chernogolovka, Moscow District, 142432, Russia}

\author{G.~Biasiol}
\affiliation{IOM CNR, Laboratorio TASC, 34149 Trieste, Italy}

\author{L.~Sorba}
\affiliation{NEST, Istituto Nanoscienze-CNR and Scuola Normale Superiore, 56127 Pisa, Italy}

\date{\today}

\begin{abstract}
We use a Mach-Zehnder quantum Hall interferometer of a novel design to investigate the interference effects at fractional filling factors. Our device brings together the advantages of usual Mach-Zehnder and Fabry-Perot quantum Hall interferometers. It realizes the simplest-for-analysis Mach-Zehnder interference scheme,  free from Coulomb blockade effects. By contrast to the standard Mach-Zehnder realization, our device does not contain an etched region inside the interference loop. For the first time for Mach-Zehnder interference scheme, the device  demonstrates interference oscillations with $\Phi^*=e/e^*\Phi_0=\Phi_0/\nu$ periodicity at fractional filling factor 1/3.  This result indicates that we  observe clear evidence for fractionally charged quasiparticles from simple Aharonov-Bohm interference. 
\end{abstract}


\maketitle

\section{Introduction}

Since the pioneering papers on quantum Hall (QH) interferometers, the fundamental goal was to extend these investigations to the regime of the fractional QH effect~\cite{stern}. Recently, interference experiments were predicted to be a powerful tool to distinguish between different proposed ground states of the $\nu=5/2$ QH liquid~\cite{feldman}. By contrast to the enigmatic $\nu=5/2$ QH liquid, there is a theoretical consensus on the properties of the primary Laughlin $\nu=1/3$ QH state. For this reason, $\nu=1/3$ filling is a good model object.

Quantum Hall interferometers are realized~\cite{stern} by means of one-dimensional transport through the current-carrying edge states (ES). ES were originally introduced~\cite{buttiker} as the intersections of the filled Landau levels with Fermi level. The local connection of two ES is equivalent to an optical semi-transparent mirror, so the sample geometry defines the interferometer scheme. 

There are two principally different schemes: (i) an electronic analog of optical Mach-Zehnder interferometer; (ii) a Fabry-Perot interference scheme. In both cases the phase difference between the interfering paths is externally controlled by variation of the magnetic flux $\Phi$ through the interferometer loop~\cite{stern}. Naive expectation of $\Phi^*=e/e^*\Phi_0=\nu \Phi_0$ periodicity is only valid    if the vacuum $\nu=1/3$  QH state within the loop is invariant~\cite{chamon}. Otherwise, only $\Phi_0=hc/e$ period can be observed.

(i) There is an etched region inside the interference loop for the standard realization of a Mach-Zehnder scheme~\cite{heiblum,heiblum+06,litvin,roulleau}.  As ensured by the general topological argumentation~\cite{BY}, only $\Phi_0=hc/e$ periodicity can be expected in this case. Experimentally, there have been no observation of the interference at fractional fillings in this geometry, possibly because of the limitations on the minimal geometrical size of the interferometer.

(ii) By contrast,   low-size Fabry-Perot interferometers clearly demonstrate interference oscillations even in the fractional regime~\cite{Camino+05,Camino+07a,Camino+07b,Ping+09,Zhang+09,ofek,McClure+12}. On the other hand, the interference pattern was shown~\cite{ofek,McClure+12} to be determined by Coulomb blockade effects~\cite{halperin-rosenow} at fractional fillings.

Here, we use a Mach-Zehnder quantum Hall interferometer of a novel design to investigate the interference effects at fractional filling factors. Our device brings together the advantages of usual Mach-Zehnder and Fabry-Perot quantum Hall interferometers. It realizes the simplest-for-analysis Mach-Zehnder interference scheme,  free from Coulomb blockade effects. By contrast  to the standard Mach-Zehnder realization, our device does not contain an etched region inside the interference loop. For the first time for Mach-Zehnder interference scheme, the device  demonstrates interference oscillations with $\Phi^*=e/e^*\Phi_0=\Phi_0/\nu$ periodicity at fractional filling factor 1/3.  This result indicates that we  observe clear evidence for fractionally charged quasiparticles from simple Aharonov-Bohm interference.

\begin{figure}
\includegraphics*[width=0.9\columnwidth]{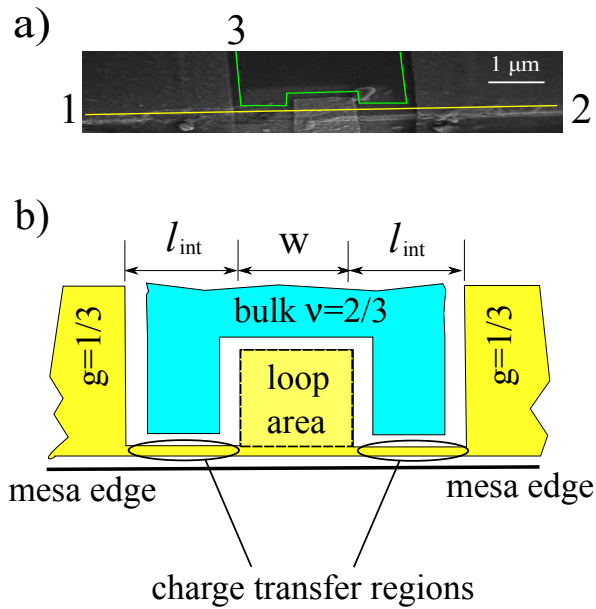}
\caption{(Color online) (a) SEM image of the interferometer. The main gate leaves the 2DEG uncovered in the narrow gate-gap region at the mesa edge. Two edge states are sketched (thick dash) for the case of filling factors $g=1$ under the gate and bulk $\nu=2$ in the gate-gap.  A small gate finger is placed at the center of this gate-gap region to detach the edge states locally. ES are independently contacted~\protect\cite{mz} by Ohmic contacts (denoted by numbers), situated far from the interferometer region.  For measurements, current is applied to the contact  1  with respect to the ground  3. In the QH regime, the current is flowing between ES to both sides of the central  gate finger. We trace the outer ES potential by the   contact 2 to study the transmittance of the gate-gap region.
(b) Schematic diagram of the compressible (white) and incompressible (color) areas of electron liquid in the interferometer region. Green (gray) color represents $\nu=2$ or 2/3 area in the gate-gap. Yellow (light gray) denotes the  $g=1$ or 1/3 QH state  under the main gate, the gate finger, and within the incompressible strip at the sample edge. Electron transport across the edge is only allowed to both sides of the central  gate finger which defines the interferometer loop area. Because of the depletion at the mesa edge, the effective  gate finger length $h$ differs~\protect\cite{mz} from the lithographic $h=0.3 \mu$m. 
\label{sample}}
\end{figure}

\section{Samples and technique}

Our samples are fabricated from a molecular beam epitaxially-grown GaAs/AlGaAs heterostructure. It contains a two-dimensional electron gas (2DEG) located 200~nm below the surface. The 2DEG mobility at 4K is  $5.5 \cdot 10^{6}  $cm$^{2}$/Vs  and the carrier density is   $1.43 \cdot 10^{11}  $cm$^{-2}$.

A novel sample design~\cite{mz} realizes a quantum Hall Mach-Zehnder interferometer based on independently contacted co-propagating edge states, see Fig.~\ref{sample} (a). The metallic gate covers the etched mesa edge, except for the narrow 3~$\mu$m region. A small gate finger is placed at the center of this gate-gap region. In a quantizing magnetic field at filling factor $\nu$ in the gate-gap region, the gate voltage is tuned to have a different QH state at filling factor $g<\nu$ under the gate.   

Let us consider the simplest case of integer $\nu=2, g=1$, see Fig.~\ref{sample} (a). There are two co-propagating ES running along the uncovered mesa edge within the gate-gap region. Because of lower filling factor $g=1$ under the gate, one (the inner) ES follows the gate edge, see Fig.~\ref{sample} (a). These ES are independently contacted, so the geometry allows a direct investigation of transport between two co-propagating ES, see Ref.~\cite{ESreview} for a review of experiments in this geometry. 

\begin{figure}
\includegraphics[width=\columnwidth]{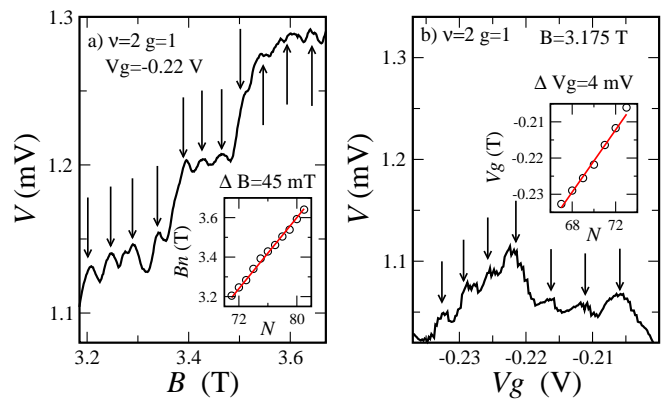}
\caption{ (Color online) Examples of the oscillating behavior while sweeping the magnetic field  $B$  at constant gate voltage (a) or vice versa (b) for the sample A at integer filling factors $\nu=2, g=1$. Insets demonstrate the positions of the oscillations (denoted by arrows in the main figures) as function of their numbers $N=B/\Delta B$. The oscillations are equidistant with periods $\Delta B=45$~mT (a) and $\Delta V_{g}=4$~mV (b) respectively. Measurement current is $I=10$~nA.    \label{oscill21}}
\end{figure}

The gate finger at the center of the  gate-gap region divides the ES junction into two ones, see Fig.~\ref{sample} (a). A particular electron can either be transferred between ES in the first ES junction of width $l_{int}$ or it can encircle the gate finger and be transferred in the second junction. If the transfer process preserves the coherence, the interference between these two trajectories should contribute to the transmittance of the device. In this case, two transmission regions serve as two semi-transparent mirrors in optical Mach-Zehnder interferometer, while two paths around the gate finger define the interferometer arms. 

The phase difference  between the interferometer arms is controlled through the Aharonov-Bohm phase $\phi=2\pi\Phi/\Phi_0$, where $\Phi$ is a magnetic flux encircled by two ES in the gate finger area. $\Phi$ can be affected by low variation either  of the magnetic field $B$ or  the effective gate finger area $S$ through the {\em top} gate voltage $V_g$. We discuss the device operation in detail below, after presenting the experimental results.

We study samples with two  gate finger widths $w=$1.5~$\mu$m (A) or 1~$\mu$m (B) with  different $l_{int}=0.75 \mu$m or 1~$\mu$m, respectively. The measurements are performed in a dilution refrigerator with the minimal temperature of 30~mK. The interference pattern is independent of the cooling cycle. Standard two-point magnetoresistance is used to determine the regions of $B$ which correspond to QH states in the ungated area.  Magnetocapacitance allows to find $V_g$ regions of QH states under the gate.

\section{Experimental results}

To measure the transmittance of the device at filling factors $\nu,g$,   dc current $I$ is applied to the outer contact 1 with respect to the grounded inner contact 3, see Fig.~\ref{sample} (a). The outer contact 2 is used to trace the outer ES potential $V$ at the other side of the gate-gap junction, i.e. it reflects the transmittance of the device~\cite{ESreview}.

To study the interference effects in the transmittance of the device, we fix the dc current $I$ and vary the magnetic field $B$ at fixed gate voltage $V_g$, see Fig.~\ref{oscill21} (a), or vice versa (b). Both $B,V_g$ are varied strictly within the  $g=1$ QH state under the gate finger to preserve the experimental geometry.

The dependencies $V(B)$, $V(V_g)$  exhibit rapid oscillations of the same amplitude against a smooth background.  They are nearly equidistant, see insets to Fig.~\ref{oscill21} (a) and (b), with periods $\Delta B=45$~mT and $\Delta V_{g}=4$~mV for the sample A. The sample B with smaller $w=1.0 \mu$m  demonstrates~\cite{mz} higher periods $\Delta B=67$~mT, $\Delta V_{g}=8$~mV. Thus, the oscillations definitely originate from the gate finger region and their period $\Delta B$ scales with the gate finger dimensions~\cite{mz}.

\begin{figure}
\includegraphics[width=\columnwidth]{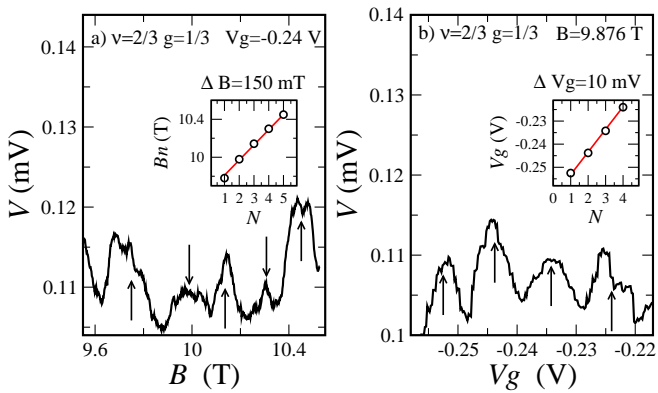}%
\caption{ (Color online) Examples of the oscillating behavior while sweeping the magnetic field  $B$  at constant gate voltage (a) or vice versa (b) for the sample A at fractional filling factors $\nu=2/3, g=1/3$. Insets demonstrate the positions of the oscillations (denoted by arrows in the main figures) as function of successive numbers $N$. The oscillations are equidistant with periods $\Delta B=150$~mT (a) and $\Delta V_{g}=10$~mV (b) respectively. Measurement current is $I=0.3$~nA.    \label{oscill2313}}
\end{figure}

\begin{figure}
\includegraphics[width=\columnwidth]{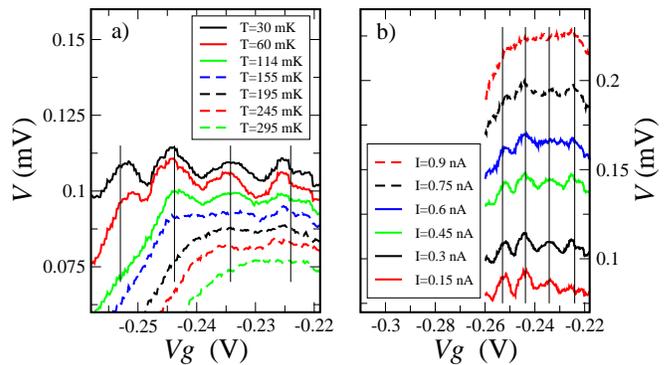}
\caption{ (Color online) Damping of the oscillations by increasing the bath temperature (a) or the imbalance (b) at fractional filling factors $\nu=2/3, g=1/3$ for the sample A, $B=9.876$~T. Measurement current is $I=0.3$~nA for the curves in the panel (a). The bath temperature is $T=30$~mK      for the curves in the panel (b). \label{amp2313}}
\end{figure}

\begin{figure}
\includegraphics[width=\columnwidth]{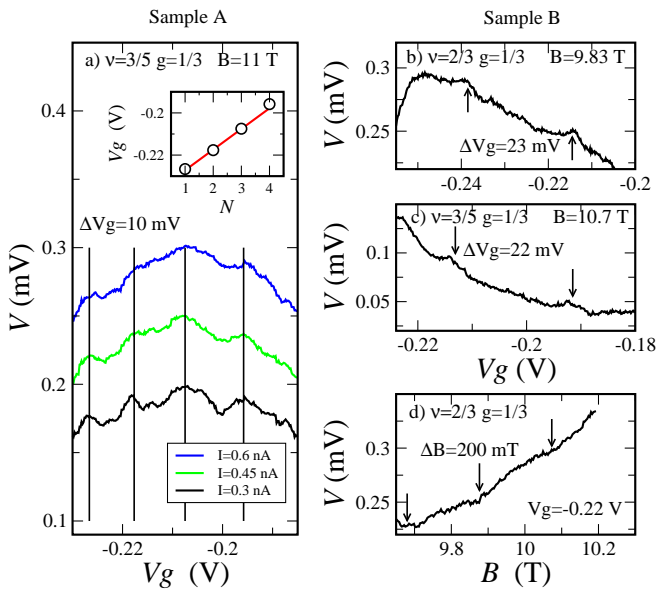}
\caption{ (Color online) (a) Damping of the oscillations by increasing the imbalance  at fractional filling factors $\nu=3/5, g=1/3$ for the sample A.   (b,c,d) Shallow oscillations at $g=1/3$ under the gate finger for the sample B at the bulk $\nu=2/3,3/5$: sweeping the gate voltage  $V_g$  at constant magnetic field (b,c); sweeping the magnetic field  $B$ at constant gate voltage (d). Measurement currents are $I=1$~nA in (b,d), and $I=0.3$~nA in (c). The bath temperature is $T=30$~mK in (a-d).    \label{oscill3513}}
\end{figure}

Fig.~\ref{oscill2313} demonstrates the oscillations in the fractional QH regime at filling factors $\nu=2/3, g=1/3$ for the sample A. They are of much smaller amplitude and are characterized by higher periods  $\Delta B=150$~mT and $\Delta V_{g}=10$~mV. The oscillations' amplitude is very sensitive to the bath temperature, see Fig.~\ref{amp2313} (a),  and to the applied current $I$, see Fig.~\ref{amp2313} (b). They disappears above $T=0.15 K$  or above $I=0.75$~nA. Both the temperature and the imbalance do  not affect the phase of the oscillations. Oscillations with the same $\Delta V_{g}=10$~mV can also be seen  at the same $g=1/3$ under the gate finger but at another bulk $\nu=3/5$, see Fig.~\ref{oscill3513} (a). 

For the sample B with narrower gate finger, oscillations are much less pronounced at fractional fillings possibly because of  wider interaction regions $l_{int}=1 \mu$m, see Fig.~\ref{oscill3513} (b,c,d). Despite the lower visibility and higher periods, these oscillations support the experimental finding $\Delta B^{[1/3]}\approx 3\Delta B^{[1]}$ and  $\Delta V^{[1/3]}_{g}\approx 3 \Delta V^{[1]}_{g} $ for the interference at fractional filling factors in Mach-Zehnder geometry.

\section{Discussion}

Low-size interferometers~\cite{Camino+05,Camino+07a,Camino+07b,Ping+09,Zhang+09,ofek,McClure+12} are   subject~\cite{ofek,McClure+12} to Coulomb blockade effects~\cite{halperin-rosenow} because the size-induced Coulomb gap becomes comparable with the corresponding spectrum  (cyclotron, Zeeman or fractional) one. 

By contrast, our device was demonstrated~\cite{mz} to operate in the extreme Aharonov-Bohm (AB) regime, free from Coulomb blockade effects, at integer fillings, i.e.  the interference period corresponds to the change of the flux $\Phi=BS$ through the interferometer loop area $S$ by one flux quantum $\Phi_0$. We can expect the same AB regime also at $g=1/3$ for the same size interferometer, since both energy scales are roughly kept constant: the fractional gap at 1/3 is of the order of the Zeeman one in 3 times lower magnetic field~\cite{vadik}.

To understand the origin of the extreme AB regime, the above edge state picture should be reformulated in terms of compressible/incompressible stripes of electron liquid~\cite{shklovskii} at the etched mesa edge. In the simplest situation of $\nu=2$ or 2/3 bulk QH incompressible state in the gate-gap,  the compressible region at the mesa edge is divided into two by a single narrow incompressible stripe  with local filling factor $\nu_c=1$ or 1/3, respectively, see Fig.~\ref{sample} (b). We deplete the 2DEG under the gate to the same filling factor $g=\nu_c$, so these compressible stripes are separated by an entire incompressible state. The geometry therefore allows  intra-edge transport investigations~\cite{ESreview}: because of macroscopic gate dimensions, charge transfer takes place only in the gate-gap junction across the $\nu_c=1/3$ incompressible stripe. Gate finger expands the stripe's width underneath, locally damping the intra-edge transport, see Fig.~\ref{sample} (b). 

 The specifics of the transport in the presented geometry is the origin of the extreme AB regime even for low-size interferometers. The standard Fabry-Perot interferometer represents a conducting island surrounded by the incompressible state of lower filling factor. To observe the interference, an electron should be added to the island, so the interior of the interferometer loop is subjected to charging~\cite{halperin-rosenow}. By contrast, in our Mach-Zehnder geometry, charge transfer takes place on both sides of the central incompressible island, determined by the  QH liquid under the gate finger, see Fig.~\ref{sample} (b).  Thus, charge transfer through the device is not connected with charging of the interior of the interferometer loop.

The present device, therefore, brings together the advantages of usual Mach-Zehnder and Fabry-Perot quantum Hall interferometers. It operates at fractional fillings in the simplest-for-analysis extreme AB regime,  free from Coulomb blockade effects,  and  the interferometer loop contains nothing but quantum Hall liquid  (there is no any etched region inside). 

Following Ref.~\cite{chamon}, even in this simplest-for-analysis regime, one should be careful about considering changes in the bulk of the quantum Hall fluid as the flux is varied: (i) $\Delta \Phi= \Delta (BS)=\Phi_0$ if the excitations are created within the loop, or (ii) $\Delta \Phi= \Delta (BS)=\Phi_0^*=e/e^*\Phi_0=3\Phi_0$ if the vacuum 1/3 QH state within the loop is invariant, i.e. quasiparticles are not formed in the bulk of the quantum Hall liquid within the loop.

From the experimental point of view,  (i) if one changes the magnetic field within the QH state at fixed gate voltage,  both possibilities of Ref.~\cite{chamon} could be imagined, because the interferometer loop is formed locally  at the edge of the macroscopic sample in our device. Thus, we should be careful in the interpretation of the experimental results in this regime.  (ii) if one changes the gate voltage within the QH state  at fixed magnetic field, the concentration is obviously constant, because the QH liquid is incompressible at the QH plateaus~\cite{aristov}. This is the regime of invariant vacuum QH state within the loop.  

Let us start from the regime of the magnetic field sweep  at fixed gate voltage. It is worth mentioning, that the extreme Aharonov-Bohm  regime means the simplest relation $\Phi=BS$, so $\Delta \Phi=S\Delta B$.  It seems to be quite natural to ascribe the difference between $\Delta B^{[1/3]}$ and  $\Delta B^{[1]}$ to the change in the effective loop area $S$. However, the relation  $\Delta B^{[1/3]}\approx 3\Delta B^{[1]}$  corresponds to the $S$ change in 3 times, which (i) seems to be too high; (ii) is very close to the filling factors ratio; (iii) is well reproducible for two different samples (A and B), so it hardly can occur incidentally. Moreover, in the integer QH regime $S$ was demonstrated to be independent of the magnetic field and integer filling factors~\cite{mz}. 

From these arguments, we have to assume that the effective loop area $S$ is roughly constant. Thus, the experimental relation $\Delta B^{[1/3]}\approx 3\Delta B^{[1]}$  indicates $\Phi^*=3\Phi_0=e/e^*\Phi_0$ flux periodicity for our Mach-Zehnder interferometer at the fractional filling factor 1/3.

In the regime of fixed magnetic field, we can only expect $\Phi^*=e/e^*\Phi_0$ flux periodicity. The analysis of the experiment is however not so straightforward, because  one should connect the experimentally observed $\Delta V_g$ with the corresponding $\Delta S$.

Variation $\delta V_g$ slightly changes the effective interferometer area $S$ because of electrostatic varying the effective gate finger perimeter. In the simplest capacitor model $e\delta N = C \delta V_g$, where $C$ is the capacitance between the gate and the compressible region around the gate finger, $\delta N$ is the variation of the compressible stripe charge,  proportional to $\delta S$ and the Landau level degeneracy $\frac{B}{\Phi_0}$. We easily obtain $\delta V_g \sim  \frac{e}{C} \frac{B}{\Phi_0} \delta S$ (cp. Ref~\cite{ofek} where a similar relation is obtained for the opposite Coulomb-dominated regime; the difference is because the AB regime is not connected with the charging of the loop area). 

Thus, the flux change in the extreme Aharonov-Bohm  regime $B\Delta S$ is directly proportional to the gate voltage period $B\Delta S \sim \frac{C\Phi_0}{e} \Delta V_g$. This relation is well fulfilled in the integer regime~\cite{mz}. The roughly constant effective loop area $S$ means the roughly constant geometric capacitance $C$.  The experimental relation $\Delta V^{[1/3]}_{g}\approx 3 \Delta V^{[1]}_{g} $ therefore indicates $\Phi^*=3\Phi_0=e/e^*\Phi_0$ flux periodicity for our Mach-Zehnder interferometer in this regime also.

As a result, our Mach-Zehnder interferometer for the first time demonstrates interference oscillations at the fractional filling factor 1/3. The observed interference corresponds to the extreme Aharonov-Bohm regime with  $\Phi^*=3\Phi_0=e/e^*\Phi_0$ flux periodicity in both regimes of the magnetic flux variation through the interferometer loop.

\acknowledgments

We wish to thank  V.T.~Dolgopolov and D.E. Feldman for fruitful discussions, and A.~Ganczarczyk for the fabrication of the sample B.
We gratefully acknowledge financial support by the RFBR and RAS.

\end{document}